\documentclass[twocolumn,aps,pre,showpacs]{revtex4}
\usepackage{graphicx}
\usepackage{amsmath}

\begin{document}

\title{A New Simulated Annealing Algorithm for the Multiple Sequence Alignment Problem: The approach of Polymers in a Random Media}

\author{M. Hern\'andez-Gu\'{\i}a}
\affiliation{Henri-Poincar\'e Group of Complex Systems, 
Physics Faculty, University of Havana, La Habana, CP 10400, Cuba}
\affiliation{
National Bioinformatics Center, Industria y San Jos\'e, Habana Vieja, Capitolio Nacional, CP 10200, Cuba}

\author{R. Mulet}
\affiliation{Henri-Poincar\'e Group of Complex Systems, 
Physics Faculty, University of Havana, La Habana, CP 10400, Cuba}
\affiliation{
Department of Theoretical Physics, Physics Faculty, University of 
Havana, La Habana, CP 10400, Cuba}

\author{S. Rodr\'{\i}guez-P\'erez}
\affiliation{Henri-Poincar\'e Group of Complex Systems, 
Physics Faculty, University of Havana, La Habana, CP 10400, Cuba}
\affiliation{Department of Informatics, University Center of Las Tunas {\em Vladimir Illich Lenin},
Las Tunas, CP 75200, Cuba
}
\date{\today}

\begin{abstract}

We proposed a probabilistic algorithm to solve the Multiple Sequence Alignment problem. The algorithm is a Simulated Annealing (SA)  that exploits the representation of the Multiple Alignment between $D$ sequences as a directed polymer in $D$ dimensions. 
Within this representation we can easily track the evolution in the configuration space of the alignment through local moves of low computational cost.
At variance with other probabilistic algorithms proposed to solve this problem, our approach allows for the creation and deletion of gaps without extra computational cost.
The algorithm was tested aligning proteins from the kinases family.
When $D=3$ the results are consistent with those obtained using a complete algorithm. For $D>3$ where the complete algorithm fails, we show that our algorithm still converges to reasonable alignments. Moreover, we study 
the space of solutions obtained and show that depending on the number of sequences aligned the solutions are organized in different ways, suggesting a possible source of errors for progressive algorithms. 

\end{abstract}

\pacs{87.10.+e,87.15.cc,87.14.-g}

\maketitle

\section{Introduction}

The Multiple Sequence Alignment (MSA) problem constitutes one of the 
fundamental 
research areas in Bioinformatics. While at first sight it may seem a simple extension of the two-string alignment problem {\em two strings good, four strings better}, for biologists, the multiple alignment of proteins or DNA is crucial in deducing their common properties \cite{Gunsfield}. Quoting Arthur Lensk \cite{Arthur,Gunsfield}: {\em One or two homologous sequences whisper... a full multiple alignment shouts out loud}. 

In general, the sequences consist of a linear array of symbols from an alphabet of $k$-letters ($k=4$ for DNA and $k=20$ for proteins).  Given $D$ sequences to determine a good Multiple Sequence Alignment is a relative task. Usually one defines a score function that depends on the distances between the letters of the alphabet, and assumes that the better alignment is the one that minimizes this score function. 

It is a common use to define the MSA score in terms of the scores of the pairwise global alignments of the sequences (Sum of Pairs score)\cite{Waterman}. 
Given two sequences $\vec{a}=a_1 \dots a_m$ and $\vec{b}=a_1 \dots b_m$ let $\Delta(a,b)$ be a cost of the mutation of $a$ into $b$  and $\gamma$ the cost of inserting or deleting of the letter. Extending  $\Delta(a,b)$ so that $\Delta(a,-)=\gamma$ and $\Delta(-,b)=\gamma$ and considering that a null (-) symbol isolated from others (-) pays an extra cost $\delta$ \cite{Waterman}
 we may define the score of a pairwise alignment $M_{i,j}$ for sequences $a_{i}$ and $b_{j}$ of size $m$ as:
 
\begin{equation}
\label{eq:sum_of_pairs}
s(M_{i,j})=\sum_{h=1}^m \Delta(a_{i,h},b_{j,h})+n \delta
\end{equation}

\noindent where $n$ is the number of isolated (-). Then,  the score for the multiple alignment \cal{M} is given by:

\begin{equation}
\label{eq:cost_function}
E(M_{i,j})=\sum_{i,j} s(M_{i,j})
\end{equation}

The multiple sequence alignment has at least three important applications in Biology: classification of protein families and superfamilies, the identification and representation of conserved sequences features of both DNA or proteins that correlate structure and/or function and the deduction of the evolutionary history of the sequences studied \cite{Gunsfield, Prevzner}.

Unfortunately the problem is known to be {\em NP-complete} and no complete algorithm exist to solve real or random instances. Therefore,  many heuristic algorithms have been proposed to solve this problem. The algorithm of Carrillo-Lipman \cite{C-L} (which is complete), is a Dynamic Programming algorithm able to find the multiple alignment of 3 sequences, and
 with some heuristic added, to find the alignments, in reasonable time,
of up to 6 sequences \cite{Gunsfield}. However, its computational cost scales very fast with the number of sequences and is of little utility for more ambitious tasks. In the first 90's the problem was approached using ideas coming from physics, J. Kim and collaborators \cite{Kim} and M. Ishikawa and collaborators \cite{Ishikawa} used different version of the Simulated Annealing \cite{Kirkpatrick} technique with some success, but their algorithms were unable to change the number of gaps in the alignment. This means that once they started with a given initial configuration (usually taken from some heuristics), any motion of segments in the sequences conserved the number of gaps. To extend these programs allowing the number of gaps to change will cause the appearance of global moves in the algorithm that are very
expensive from the computational point of view. 

Probably the must successful attempt to solve this problem has been the Clustal project \cite{paper_marylens}, a progressive algorithm that first organizes the sequences according to their distances and then
 aligns the sequences in a progressive way, starting with the most related ones. Moreover, it uses a lot of biological information, some motifs of residues rarely accept gaps, sub-sequences of residues associated with structural sub-units are preferred to stay together during the alignment, etc. These features, and a platform easy to use and integrated with other standard bioinformatic tools, have made Clustal the favorite Multiple Sequence Alignment program for biologists and people doing bionformatics in general \cite{our_note}. However, it also has important drawbacks. Once the first $k$ sequences are aligned, the inclusion of a new sequence would not change the previous alignment, 
the gap penalties are the same independently on how many sequences have been already aligned or their properties, and being a progressive method the global minimum obtained is strongly biased by those sequences which are more similar\cite{paper_marylens}.

Another, recent and also successful approach uses the concepts of Hidden Markov Models \cite{HMM}. While some of the previous drawback associated to Clustal disappear, because for example, the sequences do not need to be organized a priori, one most start assuming a known model of protein (or DNA) organization, which is usually obtained after training the program in a subset of sequences. Then, one must be aware that the results usually depend on the training set, specially if it is not too large. Moreover, if we are dealing with sequences of unknown family, or difficult to be characterized this approach does not guarantee good alignments.

Therefore we decided to propose a new Simulated Annealing (SA) algorithm that avoids the main difficulty of the previous attempts \cite{Kim, Ishikawa}. Our algorithm allows for the insertion and deletion of gaps in a dynamic way using only local moves. It makes use of the mathematical mapping between the Multiple Sequence Alignment and a Directed Polymer in a Random Media (DPRM) pointed out some years ago by Hwa et al \cite{Hwa}. 
In such a way, it should be also possible to extrapolate all the computer facilities and techniques, developed in the field of polymers to this biological problem. 

The rest of the paper is organized in the following way. In the next section we make a short review of the theoretical foundations of our algorithm. Then in section \ref{sec:Alg} we explain the implementation details to discuss the results in section \ref{sec:Res}. Finally the conclusions are presented including and outlook for future improvements of this program.

\section{Theoretical background}
\label{sec:RS}

\begin{figure}[htb]
\includegraphics[width=0.94 \columnwidth]{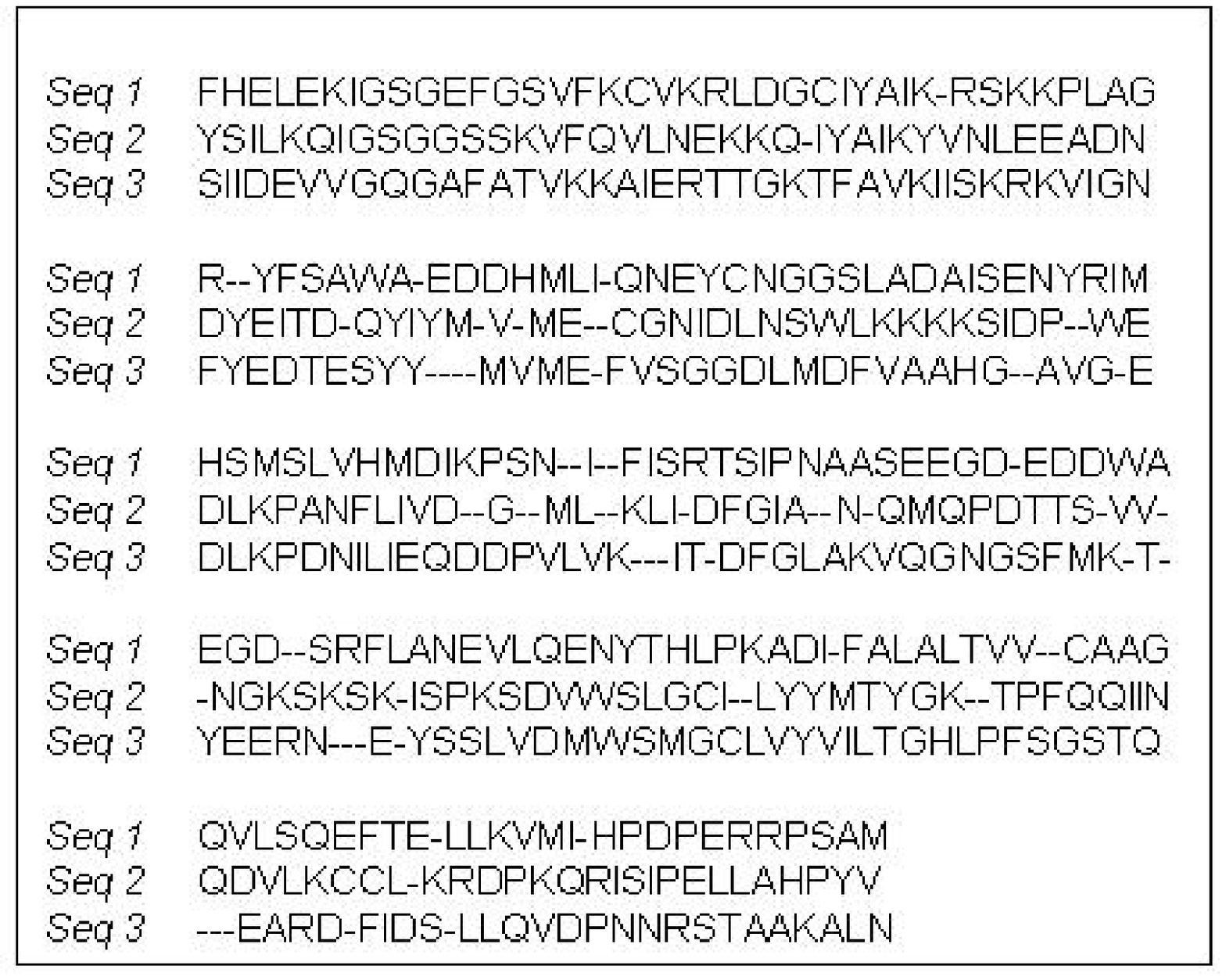}
\caption{Usual representation of a multiple sequence alignment}
\label{fig:usual_alignment}
\end{figure}

Usually, multiple sequence alignments are studied and visualized writing one sequence on the top of the other, miming a table, (see figure \ref{fig:usual_alignment})
and all the probabilistic algorithms devised so far use the simplicity of this representation to generate the moves. 

\begin{figure}[htb]
\includegraphics[width=0.94 \columnwidth]{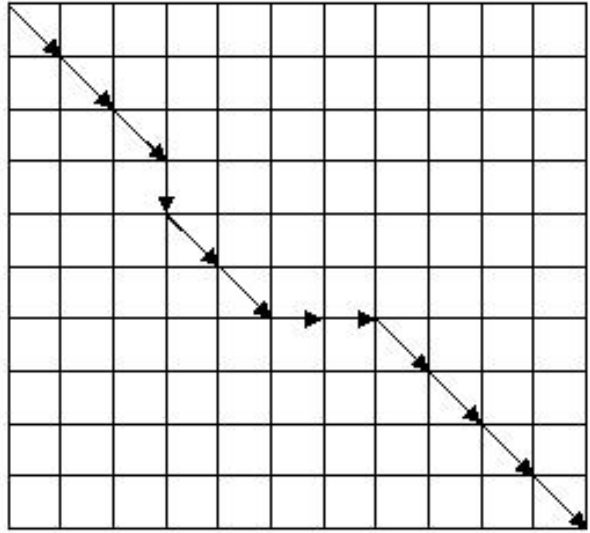}
\caption{A directed path (thick line) in a bi-dimensional grid}
\label{fig:square}
\end{figure}

Instead of that, we will use the
 well known fact \cite{Needleman,Waterman} that the alignment of $D$ sequences 
may be represented in a $D$ dimensional lattice (see figure \ref{fig:square} for $D=2$). 

The cells of the $D$-dimensional lattice are labeled by the $D$ indexes $(i_1,i_1 \ldots i_D)$. The bonds encode the adjacency of letters: A diagonal bond in a $D$ dimensional space represents the $D$-pairing $(a_{i_1},b_{i_2},\ldots,W_{i_D})$. The insertion of gaps are represented by bonds without components in the sequences where the gaps were inserted. For example, a $D$-pairing 
$(a_{i_1},b_{i_2},-,d_{i_4},\ldots,W_{i_D})$ is represented by a bond whose projection on the third sequence is zero, and the $D$-pairing $(a_{i_1},-,-,d_{i_4},\ldots,W_{i_D})$ is represented by a bond whose projection on the second and third sequences are zero.

Then, any alignment maps onto a lattice path that is directed along the diagonal of the $D$-dimensional hypercube. This lattice path may be interpreted as a Directed Polymer and the Random Media in the problem is provided by the structure of the sequences to be aligned and by the distance between the residues in the different sequences. 

This mapping was already fruitfully used by Hwa and co.\cite{Hwa} to prove that the similarities between two sequences can be detected only if their amount exceeds a threshold value and for proposing a dynamic way to determine the optimal parameters for a good alignment of two sequences. 

Here, our main focus will be to optimize the Directed Polymer (lattice path) under the constraints imposed by the sequences and their interactions in dimensions larger than 2. To use a Simulated Annealing algorithm we extend the usual representation of computer science of determining the ground state of the problem to a finite temperature description. Then, a finite-temperature alignment is a probability distribution

\begin{equation}
\label{eq:prob_function}
P(C)=\frac{1}{Z} e^{-\beta E(C)}
\end{equation}

\noindent over all possible conformations $C$ of the polymer and 
where $E$ is given by equation \ref{eq:cost_function}, and $Z$ is the partition function of the alignment \cite{Zhang}. The temperature ($\beta^{-1}$) controls the relative weight of alignments with different scores (different conformations of the polymer) while $\Delta(a,b)$ the length of the polymer and the frequency of the gaps. In physical terms, $P(C)$ defines and ensemble at temperature $\beta^{-1}$ with line tension $\gamma$ and chemical potential $\Delta(a,b)$.\cite{Lassig}

\section{The Algorithm}
\label{sec:Alg}

The Simulated Annealing (SA) was introduced many years ago by Kirkpatrick et al \cite{Kirkpatrick} to find a global minimum of a function in combinatorial optimization problems. SA is a probabilistic approach, that in general needs a state space (the different configurations of the directed path) and a cost or energy function (\ref{eq:cost_function}) to be minimize (eq. \ref{eq:cost_function}).

Simulated Annealing generates new alignments from a current alignment by applying transition rules of acceptance. The criteria for acceptance are the following:

\begin{itemize}
\item if $\Delta E <0 $, accept the new alignment
\item if $\Delta E >0 $ accept it with probability 
$P(\Delta E)=e^{-\beta (E_{new}(C)-E_{old}(C))}$
\end{itemize}

The parameter $\beta$ controls the probability to accept a new configuration. Initially, one starts at low values of $\beta$ (high temperatures) and then increases it applying an annealing schedule. If the temperature is lowered slowly enough, it can be proved that the system reaches a global minimum \cite{Gelman}. Unfortunately it will require infinitely computational time and one usually selects the best scheduling that is possible to afford with the computational facilities at hand. Then SA is run  over many initial conditions, and one assumes that the output of minimum energy is (or is close) to the global minimum. 

In the Multiple Sequence Alignment this is also the case, but differently to what happens in other combinatorial optimization problems, here the average solution, i.e. that obtained after averaging over all the local minima's may be interesting by itself. In fact, researchers are often interested not in the particular details of the alignment, but in its robust properties, and comparing all the outputs of the SA is a way to get this information.

\begin{figure}[!htb]
\begin{center}
\includegraphics[angle=0,width=.94\columnwidth]{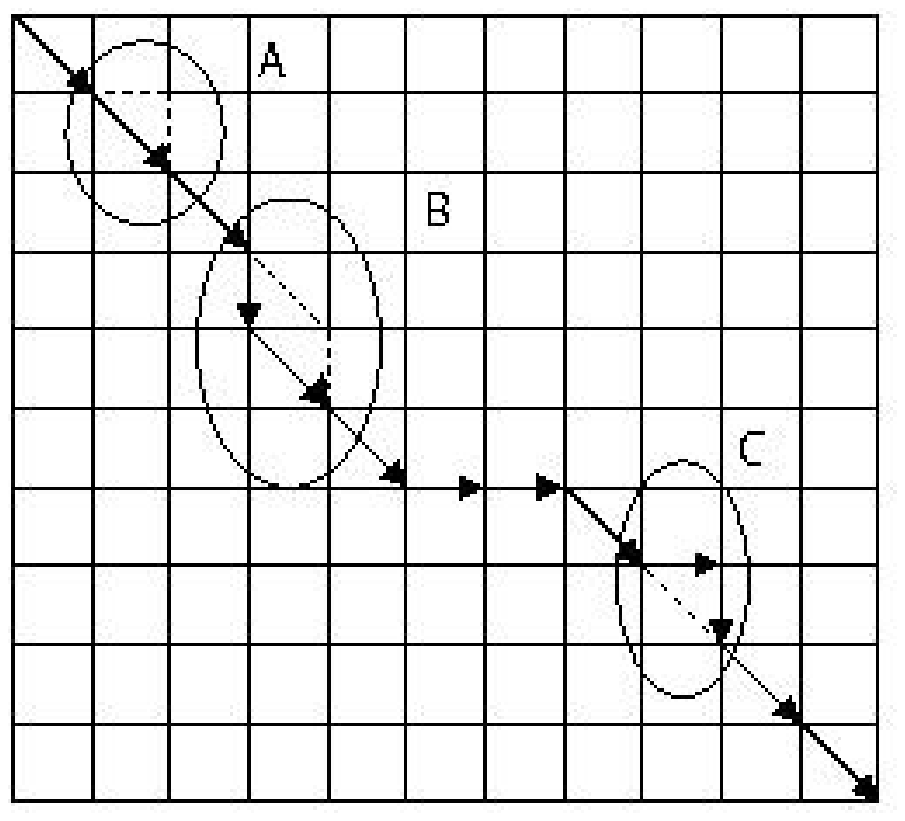}
\caption{Local moves of the algorithm in a two dimensional grid (from arrows to dashed lines), a) gap insertion, b) gap motion and c) gap deletion}
\label{fig:local_moves}
\end{center}
\end{figure}

From the technical point of view, once a cost function is defined, one needs to select the moves to be associated to the transition rates. Our description of the Multiple Sequence Alignment Problem as a Directed Polymer in a Random Media allows us to define 
three types of moves, insertion, deletion and motion of gaps. All these moves are represented in figure (\ref{fig:local_moves}) in a two-dimensional grid. The extension to $D$-dimensional systems is straightforward. 

In this way we get an algorithm that allows for the creation of gaps, which means a search space larger than the usually proved by similar methods. 
At the same time the algorithm is quadratic in the number of sequences. In fact, the computational cost of any move is limited by the square of the number of sequences to be aligned. 

In this work, we did not follow any heuristic strategy of optimization. Our intention was to prove the potentiality of this strategy and we kept things as simple as possible. For example, 
if we start too far from the global minimum,
the selection of local moves alone will make the algorithm to converge very slowly to it. This drawback may be overcome using very different initial conditions or trying, every some time steps, global moves that change radically the conformation of the polymer. We did not take care of this. During the simulation the three moves were chosen
with probability $1/3$. The only biological information inserted was given by the cost matrix used to align the protein sequences. We avoid the use of important and well know biological information, fixed residues, phylogenetic tree of the sequences, etc, and the program was designed without the use programming optimization tricks.

\section{Results}
\label{sec:Res}

All the results presented in this section reflects the alignments of proteins from the kinase family, but qualitatively similar results were obtained for the GPCRs (G protein-coupled receptors) and CRP (cAMP receptor protein) families. The simulations started
with $\beta=1.0$ and every  $t_{rel}$ montecarlo steps  $\beta$ 
was increased 
by a multiplicative factor of 1.01 until $\beta=3.0$. We took care and in all cases the system reached the equilibrium. The different initial conditions were chosen inserting gaps randomly in all the sequences but the larger one, such that considering these gaps at $t=0$, all the sequences were of equal length. To define the distance between the letters of the alphabet we use the PAM matrix \cite{HMM}.

\begin{figure}[!htb]
\begin{center}
\includegraphics[width=0.94 \columnwidth]{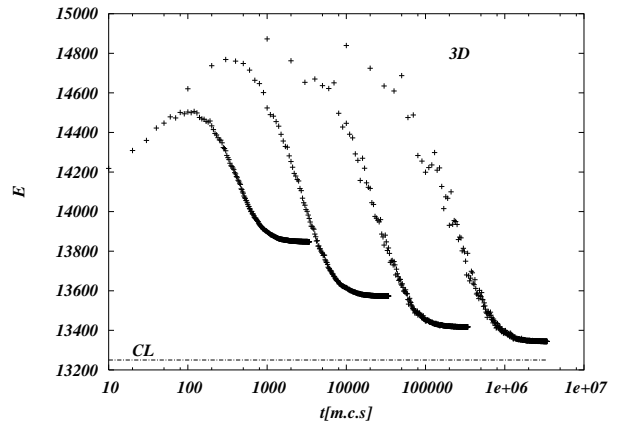}
\caption{Mean Energy versus time for the alignment of three sequences from the kinase family. From left to right: $t_{rel}=10,100,1000$ and $10000$. The average were taken over 100 initial conditions. The horizontal line represents the result of the Carrillo-Lipman algorithm}
\label{fig:relax3D}
\end{center}
\end{figure}

In figure \ref{fig:relax3D} are shown the approach to equilibrium of the multiple sequence alignment of 3 sequences averaged over 100 initial conditions for $t_{rel}=10,100,1000$ and $10000$.
Comparing with the result of  the Carrillo-Lipman 
algorithm it is evident that for $t_{rel}=10000$ the algorithm is very close to the global minimum. However, one should take care that differently to what is usually obtained in other algorithms for the Multiple Sequence Alignment problem, these figures reflects averages values of the multiple sequence alignment.

\begin{figure}[!htb]
\begin{center}
\includegraphics[width=0.94 \columnwidth]{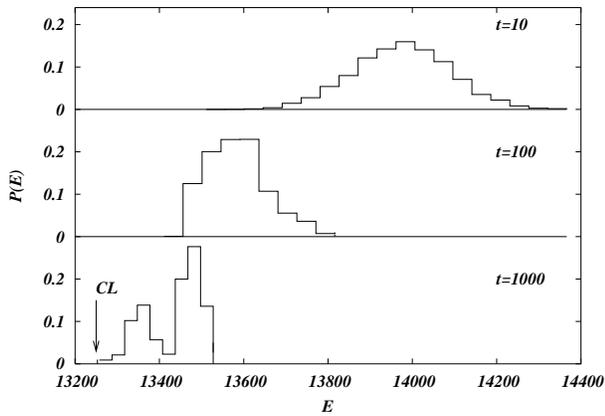}
\caption{Histograms of energy for the alignment of three sequences from the kinase family at different $t_{rel}$.}
\label{fig:hist0-3D-123}
\end{center}
\end{figure}

In fact, one is often interested, rather than on the average, in the minimum of all the alignments. In figure 
\ref{fig:hist0-3D-123} we represent an histogram in energy for the alignment, over 1000 initial conditions for different values of $t_{rel}$, of the same three proteins of the kinase family presented in figure \ref{fig:relax3D}.
Note again that for $t_{rel}=1000$ we obtain exactly the Carillo-Lipman result with probability larger than 0. Moreover, looking to the structure of the histogram for  $t_{rel}=1000$, one can also conjecture that if the average multiple sequence alignment were calculated only with those alignments concentrated in the peak of lower energies, the result presented in figure \ref{fig:relax3D} will be closer to the one of the Carrillo-Lipman algorithm.

\begin{figure}[!htb]
\begin{center}
\includegraphics[width=0.94 \columnwidth]{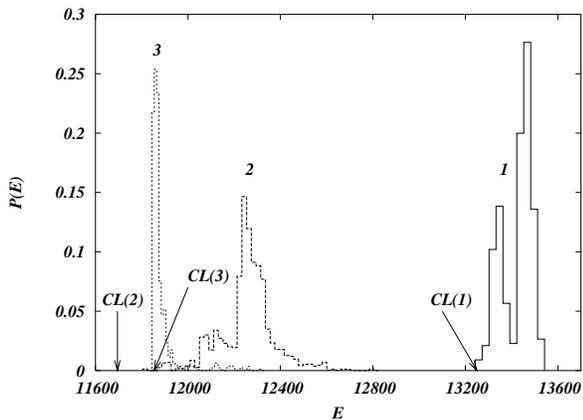}
\caption{Histograms of energy for the alignment of three different samples of three sequences from the kinase family at $t_{rel}=1000$. The Carrillo-Lipman results are represented by the arrows.}
\label{fig:hist-3D-1-9}
\end{center}
\end{figure}

Another symptom suggesting that the average over the realization must be taken with care comes from the analysis of figure \ref{fig:hist-3D-1-9}. There we present again histograms for $t_{rel}=1000$ but using three different samples of the kinase family. Note that while sample 1 and sample 3 are very well behaved and the results compare very well with the Carillo-Lipman method, the situation for sample 2 is different. To get good results in this case, it is clearly necessary to go beyond 1000 montecarlo steps.

\begin{figure}[!htb]
\begin{center}
\includegraphics[width=0.94 \columnwidth]{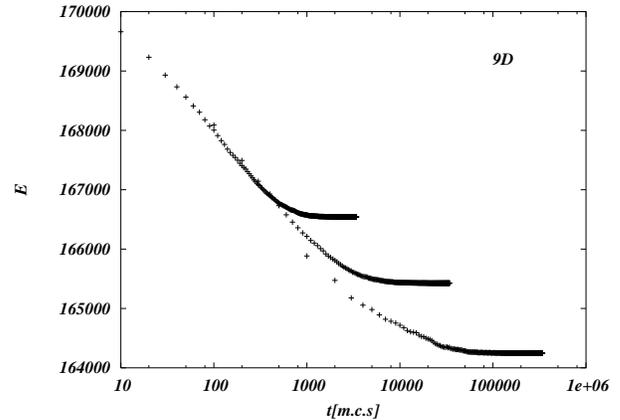}
\caption{Mean Energy versus time for the alignment of nine sequences from the kinase family. From left to right: $t_{rel}=10,100$ and $1000$. The average were taken over 100 initial conditions.}
\label{fig:relax9D}
\end{center}
\end{figure}

\begin{figure}[!htb]
\begin{center}
\includegraphics[width=0.94 \columnwidth]{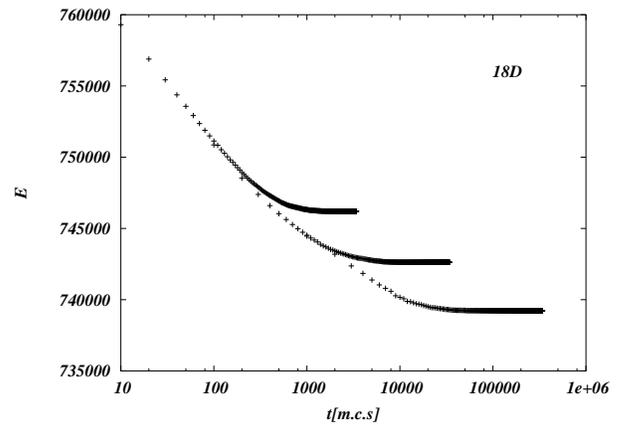}
\caption{
Mean Energy versus time for the alignment of 18 sequences from the kinase family. From left to right: $t_{rel}=10,100$ and $1000$. The average were taken over 100 initial conditions.}
\label{fig:relax18D}
\end{center}
\end{figure}

In the same spirit of figure \ref{fig:relax3D}, figures \ref{fig:relax9D} and  \ref{fig:relax18D} reflect results suggesting that also for higher dimensions, if  $t_{rel}$ is large enough the algorithm should produce good alignments. In fact, also in these cases the energy decreases linearly with $t_{rel}$.

\begin{figure}[!htb]
\begin{center}
\includegraphics[width=0.94 \columnwidth]{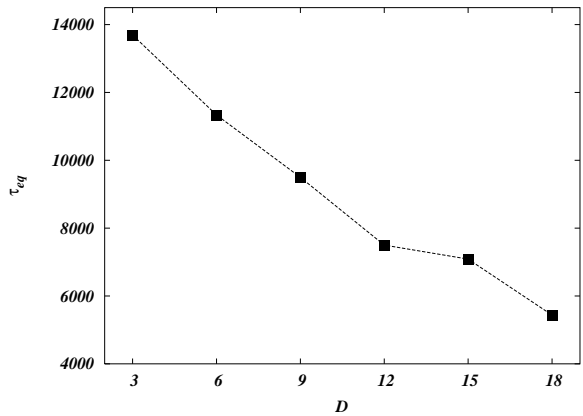}
\caption{Equilibration time versus $D$ average over 500 initial conditions for $t_{rel}=1000$.}
\label{fig:equilibritation-e-time}
\end{center}
\end{figure}

If the number of sequences is higher, the correlations between the sequences increases, and the algorithm should find better results.
This fact may be clearly seen in figure \ref{fig:equilibritation-e-time} where the equilibration time of the algorithm (in m.c.s) is shown as a function of the number of sequences. The equilibration time, measured as the time necessary to reduce the energy by  a factor $e$, 
decreases linearly with the number of sequences to be aligned.  Of course the results may change if very different sequences are aligned, but in this case all other 
known algorithms fail to predict good alignments. Then, we may say that for the most common cases, of correlated sequences, we present an algorithm whose convergence time decreases with $D$, and whose moves, only increase quadratically with $D$. 

With these results at hand, we follow to study the structure of the space of the solutions as a function of the number of aligned sequences. We define a distance ($d$) between two alignments $\cal{A}$, and $\cal{B}$ in the following way. 
Given two solutions, $A_{i,j}$ and $B_{i,j}$ (where the index $i$ stands for the sequence and $j$ for the position of the symbol in the sequence) we, aligned one by one of the $D$ sequences of each solution using a Dynamic Programming algorithm reminiscent of the Needleman-Wunsh algorithm with the following score function:

\[c_a(A,B) = \left\{ \begin{array}{ll}
		0 & \mbox{if $A_{i,j}=B_{i,j}$} \\
		1 & \mbox{if $A_{i,j} \neq B_{i,j}$} \\
		r>1 & \mbox{if a gap is inserted}
		\end{array}
	\right. \]

\noindent and express $d_{\cal{A},\cal{B}}$ as:

\begin{equation}
d_{\cal{A},\cal{B}}=\frac{1}{D} \sum_{i,j} c_a(A_{i,j},B_{i,j})
\end{equation}

In this way identical alignments will be at distance $0$ from each other, and the insertion of gaps to obtain good alignments is penalized, such that the original alignments are altered minimally during the calculation of $d_{\cal{A},\cal{B}}$. We calculated $d$ with $r=2$ and $r=8$ and the results were the same (apart a constant shift in $d$). Below we present figures for $r=2$.
 
We study how different are, from the solution of minimum score, the other solutions obtained aligning $D$ sequences. For every value of $D$ we use 1000 initial conditions and a $t_{rel}=1000$. Note that the sequences used were always the same. This mean, we first aligned three sequences from the kinase family. 
Then, to align 4 sequences we just add a new one to the previous three and started the alignment from scratch. The procedure was repeated for every new value of $D$.

\begin{figure}[!htb]
\begin{center}
\includegraphics[width=0.94 \columnwidth]{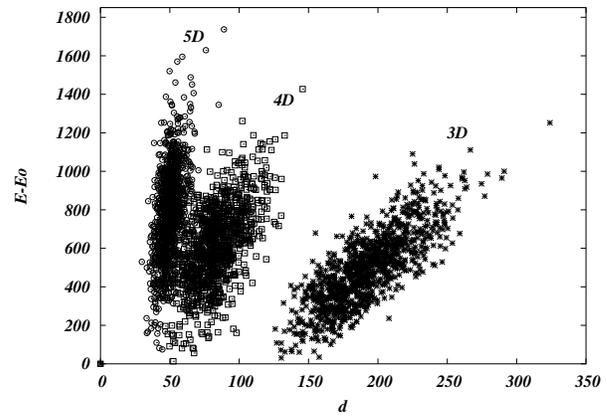}
\caption{$E-E_0$ versus $d$ for the alignment of 3,4 and 5 sequences of the kinase family}
\label{fig:clustering-result-down}
\end{center}
\end{figure}

\begin{figure}[!htb]
\begin{center}
\includegraphics[width=0.94 \columnwidth]{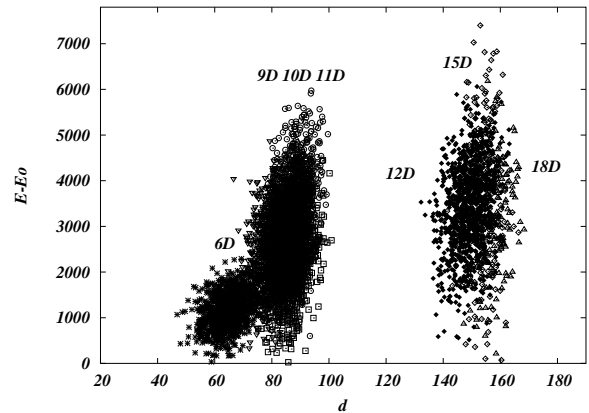}
\caption{$E-E_0$ versus $d$ for the alignment of 6,9,10,11,12,15 and 18 sequences of the kinase family}
\label{fig:clustering-result-up}
\end{center}
\end{figure}

The results appear in figures \ref{fig:clustering-result-down} and \ref{fig:clustering-result-up} where $E-E_{min}$ is plotted as a function of $d$. From the figures it becomes evident that
the space of solution strongly depends on the number of sequences aligned. 
For example, while for three sequences the distance between the alignments is correlated with the difference in score, for more than 4 sequences it is not true anymore. Moreover, while  for $D<6$ the distance between the solutions decreases, it increases for $D>6$ and remains constant for $D>12$. 
Surely, these results reflect the correlation between the proteins aligned. For example, we may speculate that for $3<D<6$, any new protein added contributed to find more similar alignments, i.e. added relevant information to the system.
However, for $D>6$ the sequences contributed with new and uncorrelated information that produce more distant alignments and 
for $D>12$ to add new proteins that belong to the kinase family will not change the relevant characteristics of the alignment. 

These results are relevant, considering the limitations of progressive algorithms to change the previous alignment when new sequences are added. Figures  \ref{fig:clustering-result-down} and \ref{fig:clustering-result-up}, clearly suggest that the inclusion of one single sequence may dramatically change the character of the solutions.

\section{Conclusions and Outlook}
\label{sec:CONC}

In this work we presented a new probabilistic algorithm, to perform the Multiple Alignment of proteins. The algorithm is based on the mapping between the DPRM and the Multiple Sequence Alignment problem. At variance with other probabilistic algorithms our algorithm permits the variation of the number of gaps in the alignment without the necessity of expensive global moves. It is proved that for small number of sequences it reproduces the results of a complete algorithm. Moreover, we show that for practical purposes the equilibration time is almost independent on the number of sequences aligned $D$ and in the worst case, it scale linearly with $D$. Finally, we study the space of solutions for different number of aligned sequences, and find a very rich structure that indicates the importance of just one sequence in the multiple alignment.

Of course the algorithm is still far from being competitive with other approaches like HMM and CLUSTAL. We are already working in implementing a similar work but using Parallel Tempering instead of Simulated Annealing. It is know that Parallel Tempering is more useful that S.A. when dealing with very hard problems, like spin glasses. Moreover, it is very suitable to parallelization. Also a direction of current work is the introduction of biological information relevant for the alignment. This may impose important constraints in the possible alignments, that may in turn strongly reduce the space of possible solutions. And last, but not least, important programing optimization are necessary to make competitive this program from the computing time point of view.

\begin{acknowledgments}
R. M. will like to thank K. Leon  and E. Moreno for useful discussion at the early stage of this work.
\end{acknowledgments}


\begin{thebibliography}{99}

\bibitem{Gunsfield} D. Gunsfield, {\em Algorithms on Strings, Trees and Sequences} (1999) Cambridge University Press.
\bibitem{Arthur} T. J. P. Hubbard, A. M. Lensk and A. Tramontano, Nature Structural Biology {\bf 4}, 313 (1996)
\bibitem{Waterman} M.S. Waterman, {\em Introduction to Computational Biology} (200), Chapm \& Hall.
\bibitem{Prevzner} P. Pevzner, {\em Computational Molecular Biology} (2000) MIT Press.
\bibitem{C-L} H. Carrillo and D. Lipman, SIAM J. Appl. Math, {\bf 48}, 1073, (1988)
\bibitem{Kim} J. Kim, S. Pramanik and M. J. Chung, CABIOS {\bf 10} 419 (1994)
\bibitem{Ishikawa} M. Ishikawa, T. Toya, M. Hoshida, K. Nitta, A. Ogiwara and M. Kanehisa, CABIOS {\bf 9} 267, (1993)
\bibitem{Kirkpatrick} S. Kirkpatrick, C.D. Gelatt and M. P. Vecchi, Science {\bf 220}, 671 (1983)
\bibitem{paper_marylens} J. D. Thompson, D. G. Higgings and T. J. Gibson, Nucleid Acids Res. {\bf 22}, 4673 (1994)
\bibitem{our_note} We are not aware of any statistical study proving this, but we believe that this statment is well accepted within the colleagues we contacted.
\bibitem{HMM} R. Durbin, S. R. Eddy, A. Krogh and G. Mitchison, {\em Biological Sequence Analysis} (1998) Cambridge University Press.
\bibitem{Hwa} T. Hwa and Michael Lassig, Phys. Rev. Lett. {\bf 76}, 2591 (1996)
\bibitem{Zhang} M. Q. Zhang and T. G. Marr, J. Theor. Biol. {\bf 174}, 119 (1995)
\bibitem{Lassig} M. Kschischo and M. L$\ddot{a}$ssig, preprint
\bibitem{Gelman} S. Geman, S. Geman, IEEE transictions on Pattern Analysis and Machine Inteligence, {\bf 6},721 (1984)
\bibitem{Needleman} S.B. Needleman and C. D. Wunsch, J. Mol. Biol. {\bf 48}, 444 (1981)
\end{thebibliography}
\end{document}